\documentclass[aps,twocolumn,amssymb]{revtex4}
\usepackage{graphicx}
\usepackage{color}
\usepackage{epsfig}
\usepackage{latexsym}
\usepackage{bm}
\usepackage{mathtools}
\newcommand{\la}{\langle}
\newcommand{\ra}{\rangle}
\newcommand{\be}{\begin{equation}}
\newcommand{\ee}{\end{equation}}
\newcommand{\bea}{\begin{eqnarray}}
\newcommand{\eea}{\end{eqnarray}}

\begin{document}

\title{Motion of active tracer in a lattice gas with cross-shaped particles}

\author{Rakesh Chatterjee$^1$, Nimrod Segall$^1$, Carl Merrigan$^2$, Kabir Ramola$^2$, Bulbul Chakraborty$^2$, and Yair Shokef$^1$}

\affiliation{$^1$School of Mechanical Engineering and Sackler Center for Computational Molecular and Materials Science, Tel Aviv University, Tel Aviv 69978, Israel \\
$^2$Martin Fisher School of Physics, Brandeis University, Waltham, MA 02454, USA}

\begin{abstract}
We analyze the dynamics of an active tracer particle embedded in a thermal lattice gas. All particles are subject to exclusion up to third nearest neighbors on the square lattice, which leads to slow dynamics at high densities. For the case with no rotational diffusion of the tracer, we derive an analytical expression for the resulting drift velocity $v$ of the tracer in terms of non-equilibrium density correlations involving the tracer particle and its neighbors, which we verify using numerical simulations. We show that the properties of the passive system alone do not adequately describe even this simple system of a single non-rotating active tracer. For large activity and low density, we develop an approximation for $v$. For the case where the tracer undergoes rotational diffusion independent of its neighbors, we relate its diffusion coefficient to the thermal diffusion coefficient and $v$. Finally we study dynamics where the rotation of the tracer is limited by the presence of neighboring particles. We find that the effect of this rotational locking may be quantitatively described in terms of a reduction of the rotation rate. 
\end{abstract}

\maketitle

\section{Introduction}

A rich variety of dynamics can occur in assemblies of particles which display independent persistent motion. Such collections, referred to as {\it active materials} are realized in several natural contexts such as proteins or motors inside cells~\cite{parry_2014, chaudhuri_2011}, monolayers of migrating cells~\cite{angelini_2011, bi_2015}, bacterial suspensions~\cite{chen_2012}, pedestrians at crowded events \cite{silverberg_2013}, and even traffic jams. At moderate densities, active particles with simple repulsive interactions can separate into inhomogeneous regions of liquid and gas, a phenomenon termed Motility-Induced Phase Separation or MIPS~\cite{fily_2012, redner_2013, cates_2015}. When confined to even higher densities, active materials can crystallize~\cite{bialke_2012}, become jammed~\cite{Liao_2018}, undergo glass transitions~\cite{Fily2014, berthier_2014, rituparno_2016}, or even exhibit gelation~\cite{redner_2013_b, levis_2014}. However, since active particles strongly perturb (and are perturbed by) the motion of the particles around them, predicting the collective dynamics of such systems is difficult. In this paper we address the simpler problem of a single active particle in a dense environment. In particular, a system that exhibits glassy dynamics in the absence of activity due to inherent geometric frustration.  

\begin{figure}[t]
\begin{center}
\includegraphics[width=\columnwidth]{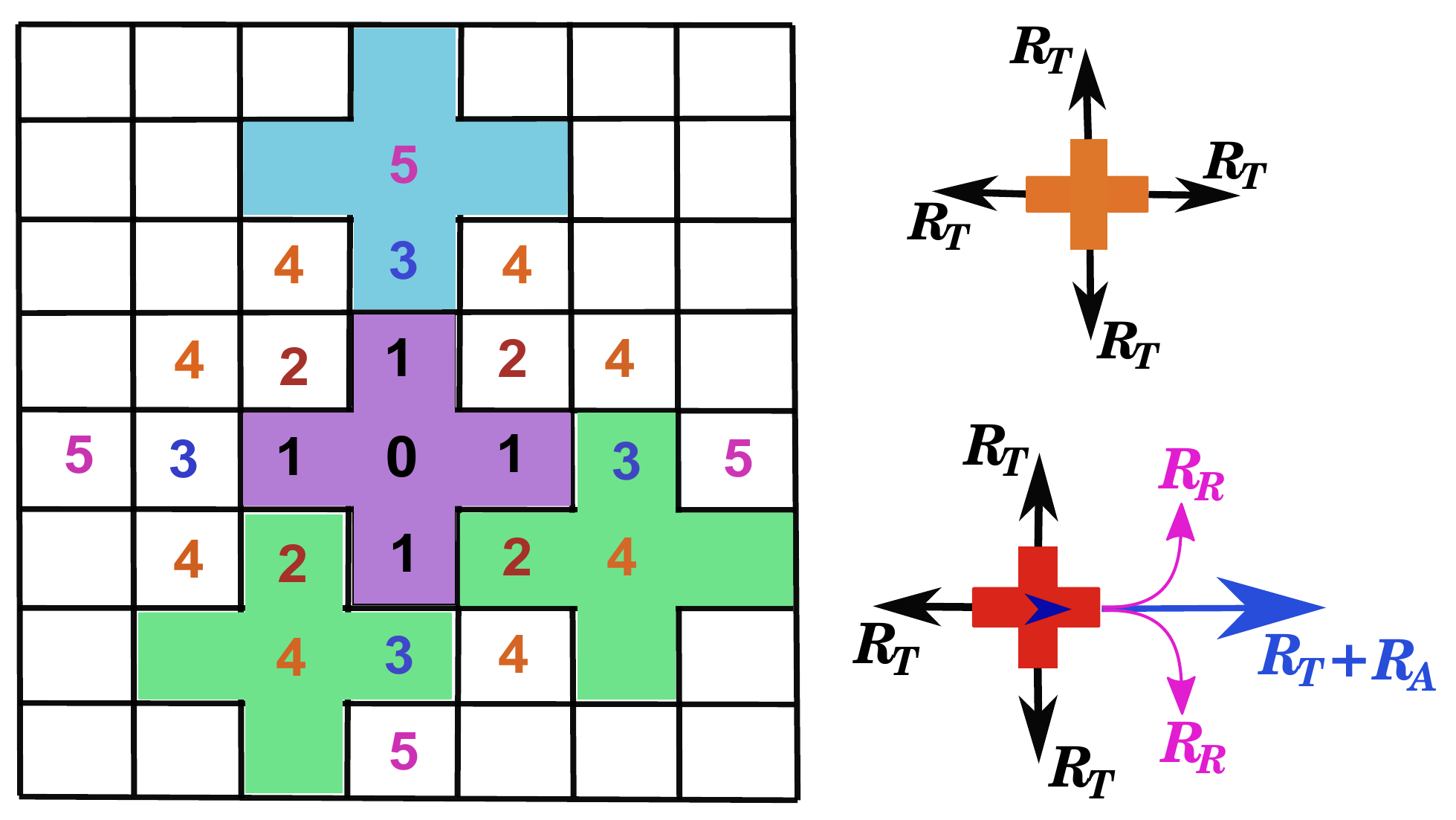}
\caption{Sites neighboring the center of a cross-shaped particle (purple) are numbered by their proximity to the center of the particle (site 0). Positions up to third nearest neighbors are excluded for centers of neighboring particles. Another particle can only sit at a fourth (green), fifth (blue) or higher-order nearest-neighbor positions. With rotational locking, fourth-order neighbors prevent rotation. Dynamics are shown for a passive particle (orange) with only thermal rate $R_T$ and for the active tracer (red) with thermal rate $R_T$, active rate $R_A$ and rotation rate $R_R$, small blue arrowhead at the center of the active tracer denotes its active direction.}
\label{fig:active-cross}
\end{center}
\end{figure}

We study a lattice gas of hard cross-shaped particles on the square lattice. Each cross prevents the occupation of its first, second, and third nearest neighbors, see Fig \ref{fig:active-cross}. We introduce a single active tracer particle, which in addition to taking thermal steps in each of the four lattice directions, also takes active steps along the direction in which it is oriented. Additionally, the tracer also performs rotational diffusion. Although thermal tracer motion has been very well studied experimentally and theoretically, much less is known about active or driven tracers. Recent studies have focused on active tracer motion in one-dimension \cite{illien_2013}, externally driven tracers in two-dimensions~\cite{Benichou2013, Benichou2014}, as well as actively moving particles near the jamming transition~\cite{Liao_2018}. Tracer motion in an embedding fluid is also an important theoretical problem in non-equilibrium statistical physics~\cite{mallick_2011, squires}.

Lattice-gas models with different sizes of the excluded-volume region around each particle have a long history of study~\cite{Bellemans1996, Baxter1980, Orban1982, Eisenberg2005, Fernandez2007, Barnes2009, Dickman2012, Nath2014, Rotman2009, Rotman2010, Eisenberg2000, Ramola2012, Barma1994}. As solid-liquid phase transitions in real fluids are mediated by strong short-ranged repulsive forces,  hard-core exclusion models are useful first approximations, and indeed exhibit many of the same features found in simple fluids, structural glasses and granular materials. The hard-cross model we study in this paper is particularly interesting in the context of dynamics. It is the simplest lattice gas model which displays a first-order phase transition to a crystal with increasing density, occurring at a melting density of $\rho_m \approx 0.16$~\cite{Fernandez2007}. Several studies have focused on the glassy behavior which results when the density of the system is quenched past this transition through random packing~\cite{Rotman2009, Rotman2010}. The glassy dynamics arise due to the ten possible close-packed sublattice orderings, which create the possibility for frustration~\cite{Eisenberg2000}. These competing, incompatible ordered structures in the same region lead to a strong suppression of the dynamics in this model. The supercooled branch is predicted to terminate at $\rho_g \approx 0.17$. We therefore expect our results to aid in the understanding of the interplay between activity and slow dynamics in frustrated systems. Finally, cross-shaped particles are also interesting since, owing to their shape, it is possible for a particle to limit the rotation of its neighbors (locking). In this paper we study the system with and without such rotational locking, and find that it has a significant effect on the tracer's dynamics. Rotational locking should have important consequences for collective effects in all active systems, which we will study in a subsequent publication.

The paper is organized as follows. In Section~\ref{sec:model} we introduce the model and describe the simulation details. In Section~\ref{sec:non-rot} we study the case of an active tracer particle that does {\it not} change orientation. We derive an analytical expression for its resulting steady-state drift velocity in terms of nonequilibrium density correlations involving the tracer and the passive particles in its vicinity. Surprisingly, we find that the properties of the passive system are not enough to fully describe the single active tracer, even in the low-activity and low-density limit. However, we do develop a theory in terms of equilibrium correlators, which is applicable in the low-density yet strong-activity regime. In Section~\ref{sec:no-lock} we study the motion of the tracer with free rotational diffusion, i.e. the active direction stochastically changes orientation independent of its neighbors. We obtain a theoretical result for the self-diffusion of the rotating active tracer in terms of the self-diffusion of a passive tracer and the drift velocity of a non-rotating tracer. Finally, in Section~\ref{sec:lock} we turn our attention to the physically relevant but much less-studied case, where particle shape affects the rotational diffusion of the tracer (rotational locking). We show that many of the results from the previous section are still applicable, with a reduced effective rotation rate compared to the rotation attempt rate. Since few studies have analyzed active or driven systems with rotational locking, our present study of a single active tracer lays the ground-work for investigation of systems with a finite density of such active particles.

\section{Model}\label{sec:model}

We study the lattice gas of particles with exclusion up to the third nearest neighbor on the square lattice. Figure~\ref{fig:active-cross} shows the equivalence of this exclusion to hard-core cross-shaped particles. Particles attempt to move to each one of their four nearest-neighbor sites at a fixed \emph{thermal rate} $R_T$. We introduce a single active tracer particle into the system which is characterized by a self-propulsion direction (North, South, East or West), along which it attempts to move with an \emph{active rate} $R_A$. This active motion of the tracer is in addition to the thermal moves in all four directions. The active particle attempts to rotate its active direction by $\pm \frac{\pi}{2}$, to either direction at a \emph{rotation rate} $R_R$, see Fig.~\ref{fig:active-cross}. In active systems the ratio of active propulsion and thermal motion is usually referred to as the P\'eclet number, thus here we identify $\textrm{Pe} = \frac{R_A}{R_T}$. Moreover, we identify the rotational diffusion coefficient with $R_R$. The dynamics of our model are equivalent to those used for the simple exclusion active lattice gas studied in~\cite{whitelam_2018}. However, as we show below, the slightly extended range of the interactions give rise to new and interesting phenomena.

Moves to a new site are accepted if they do not create overlaps between any two crosses, i.e exclusion up to third nearest neighbor. For rotation events, we consider two different versions of the dynamics. In the first version, rotations are always allowed whenever they are attempted, while in the second version crosses in fourth-order neighboring sites prevent one another from changing direction. The rotational-locking case may be interpreted in the following way: active particles that have a rigid structure must physically rotate in order for the particle to change direction. Rotational locking may also be thought of as a useful approximation for active particle which are able to exert torques. We find that analysis of the freely-rotating case is helpful in  understanding systems with rotational locking, where the particles can be described as rotating with a modified rate.

Throughout this paper we limit ourselves to the range of densities $\rho < \rho_m \approx 0.16$ that is below the first-order phase transition, hence the bath of passive particles always reaches equilibrium and does not get arrested in glassy states. In our simulations, the bath of passive particles therefore always reaches an equilibrium state. We run dynamical Monte Carlo simulations on a periodic lattice of dimension $L \times L$. We typically use $L=100$, while for very low densities below $\rho=0.01$ to get reasonable statistics with enough particles in the lattice, we increased the system size to $L=500$. We measure time in units of $R_T$. We first allow the system to relax for time $t=10^5$, and then we start measuring the drift velocity or diffusion coefficient, until $t=10^6$.

\section{Non-Rotating Active Tracer}\label{sec:non-rot}

Here we consider the zero-rotation-rate limit $R_R=0$. This is interesting in its own right, but as we will show in the following sections, this limit serves as the basis for understanding the case of a rotating active particle. We will demonstrate that the motion of a rotating active particle may be described by a decoupling between rotation events, and the persistent motion during time intervals between rotations. Thus we first need to understand the dynamics without rotation.

\subsection{Exact theory using non-equilibrium correlators}

\begin{figure}[t]
\begin{center}
\includegraphics[width=0.6\columnwidth]{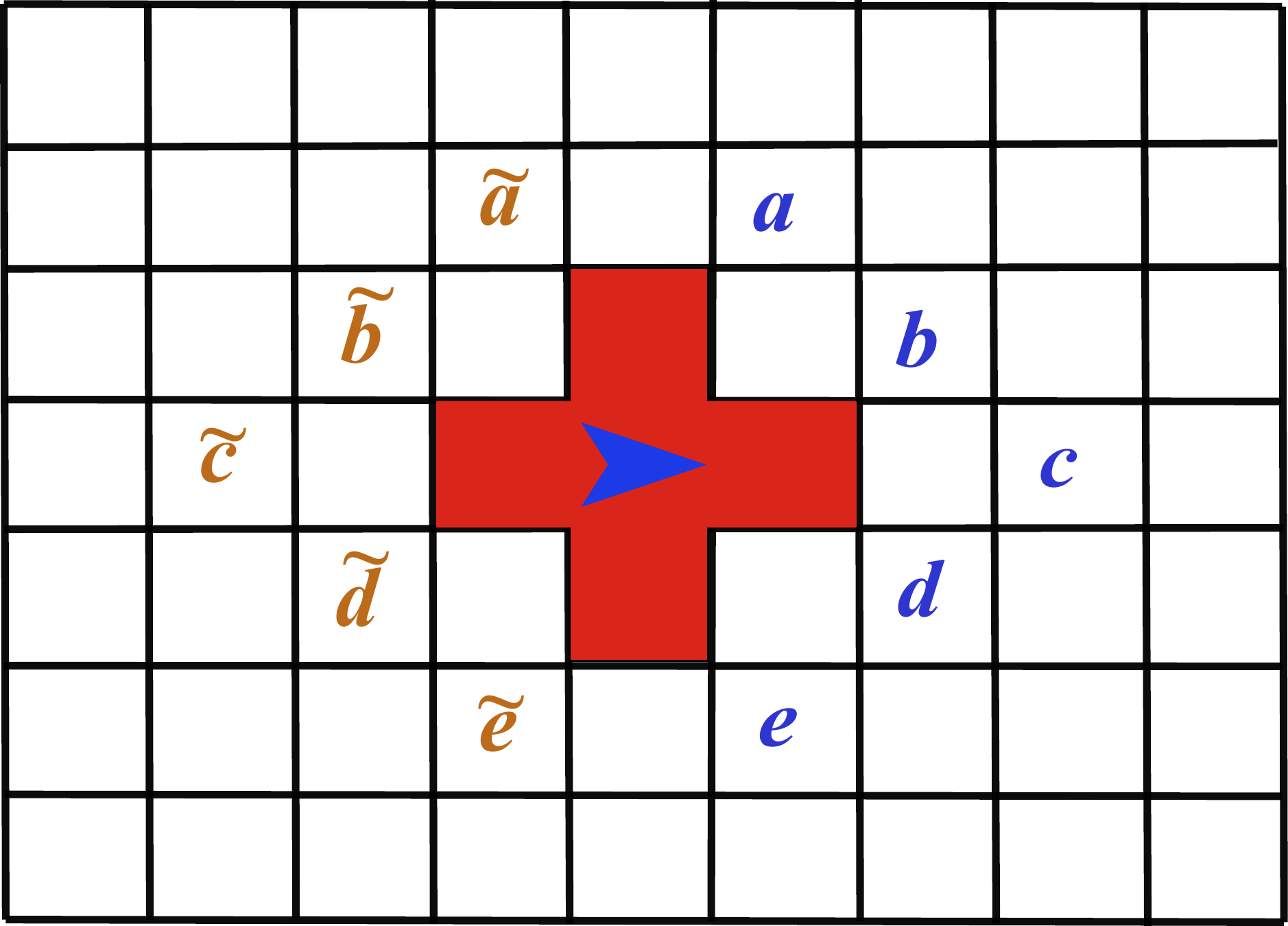}
\caption{Active tracer with its active direction denoted by a blue arrowhead. The sites, which must be vacant along the active direction are denoted by $a$, $b$, $c$, $d$, and $e$, and along the opposite direction by $\tilde{a}$, $\tilde{b}$, $\tilde{c}$, $\tilde{d}$, and $\tilde{e}$.}
\label{fig:active-nonrot}
\end{center}
\end{figure}

To study the motion of a non-rotating active tracer we note that in order for the particle to move in its active direction it needs the five sites denoted $a$, $b$, $c$, $d$, and $e$ in Fig.~\ref{fig:active-nonrot} to all be vacant. If we denote the occupancy of each site $i$ by the indicator random variable $\eta_i = 0,1$, then for a given configuration of the particles on the lattice, the ability to move in the active direction is given by
\bea
W_+=(1-\eta_a)(1-\eta_b)(1-\eta_c)(1-\eta_d)(1-\eta_e) . \label{eq:W+}
\eea
This variable may be equal to zero or one, with $W_+=1$ meaning that the particle can move forward and $W_+=0$ meaning that it cannot. To analyze the \emph{net} flow in the active direction we consider also the motion in the opposite direction, which in turn requires vacancy of all sites $\tilde{a}$, $\tilde{b}$, $\tilde{c}$, $\tilde{d}$, and $\tilde{e}$. The ability to move in the opposite direction is given by
\bea
W_-=(1-\eta_{\tilde{a}})(1-\eta_{\tilde{b}})(1-\eta_{\tilde{c}})(1-\eta_{\tilde{d}})(1-\eta_{\tilde{e}}) . \label{eq:W-}
\eea

Now, the position $r_\parallel(t)$ along the active direction of this non-rotating active tracer evolves according to the following stochastic dynamics:
\bea
r_\parallel(t+dt) = 
\begin{dcases}
 \underline{\textrm{value}}:  &  \underline{\textrm{probability:}} \\
 r_\parallel(t) + 1 &  p_+ \\
 r_\parallel(t) - 1 &  p_- \\
 r_\parallel(t)     &  1 - p_+ - p_- ,
\end{dcases}
\label{evolution_eqns}
\eea
where $p_+ = (R_T+R_A) W_+ dt$ and $p_- = R_T W_- dt$ are the probabilities to move in the forward and backward directions, respectively during an infinitesimal time interval $dt$. Averaging over Eq.~(\ref{evolution_eqns}) we see that the average drift velocity of the active tracer is given by,
\begin{eqnarray}
v \equiv \frac{d \langle r_\parallel(t) \rangle}{dt} &=& (R_A+R_T) {\cal{C}_+} - R_T {\cal{C}_-} ,
  \label{eqn:v_x_1}
\end{eqnarray}
where ${\cal{C}_{\pm}} = \langle W_{\pm} \rangle$ are the probabilities that the moves in the forward and backward directions are not blocked by other particles. Note that these probabilities depend on non-equilibrium correlations that develop in the close proximity of the active tracer due to its non-equilibrium motion, and thus depend not only on density, but also on activity.

We now expand the products in Eqs.~(\ref{eq:W+}-\ref{eq:W-}). We note that due to the model's exclusion, some blocking sites may not be simultaneously occupied, for instance sites $a$ and $b$. Therefore $\eta_a$ and $\eta_b$ may not be both equal to one, thus $\la \eta_a \eta_b \ra = 0$. By canceling all such terms that are identically zero, we may write 
\bea
{\cal{C}_+} = 
\Big( 1 &-& \la \eta_a \ra - \la \eta_b \ra - \la \eta_c \ra - \la \eta_d \ra - \la \eta_e \ra \nonumber \\ 
&+& \la \eta_a \eta_c  \ra + \la \eta_a \eta_d  \ra + \la \eta_a \eta_e  \ra + \la \eta_b \eta_e \ra + \la \eta_c \eta_e  \ra \nonumber \\ 
&-& \la \eta_a \eta_c \eta_e  \ra \Big) .
\eea
Sites $a$ and $e$ are symmetric and also $b$ and $d$, so this may be further simplified to the following form
\begin{eqnarray}
{\cal{C}_+} &=& \Big(1 - 2 \la \eta_a \ra - 2 \la \eta_b \ra - \la \eta_c \ra \nonumber \\ &+& 2 \la \eta_a \eta_c  \ra + 2 \la \eta_a \eta_d  \ra + \la \eta_a \eta_e  \ra  - \la \eta_a \eta_c \eta_e  \ra\Big) , \label{eq:C+}
\end{eqnarray}
and for the backward direction we similarly obtain
\begin{eqnarray}
{\cal{C}_-} &=& 
\Big(1 - 2\la \eta_{\tilde{a}} \ra - 2 \la \eta_{\tilde{b}} \ra - \la \eta_{\tilde{c}} \ra \nonumber \\ &+& 2 \la \eta_{\tilde{a}} \eta_{\tilde{c}} \ra + 2 \la \eta_{\tilde{a}} \eta_{\tilde{d}}  \ra + \la \eta_{\tilde{a}} \eta_{\tilde{e}} \ra - \la \eta_{\tilde{a}} \eta_{\tilde{c}} \eta_{\tilde{e}} \ra \Big) . \label{eq:C-}
\end{eqnarray}

\begin{figure}[t]
\includegraphics[width=\columnwidth]{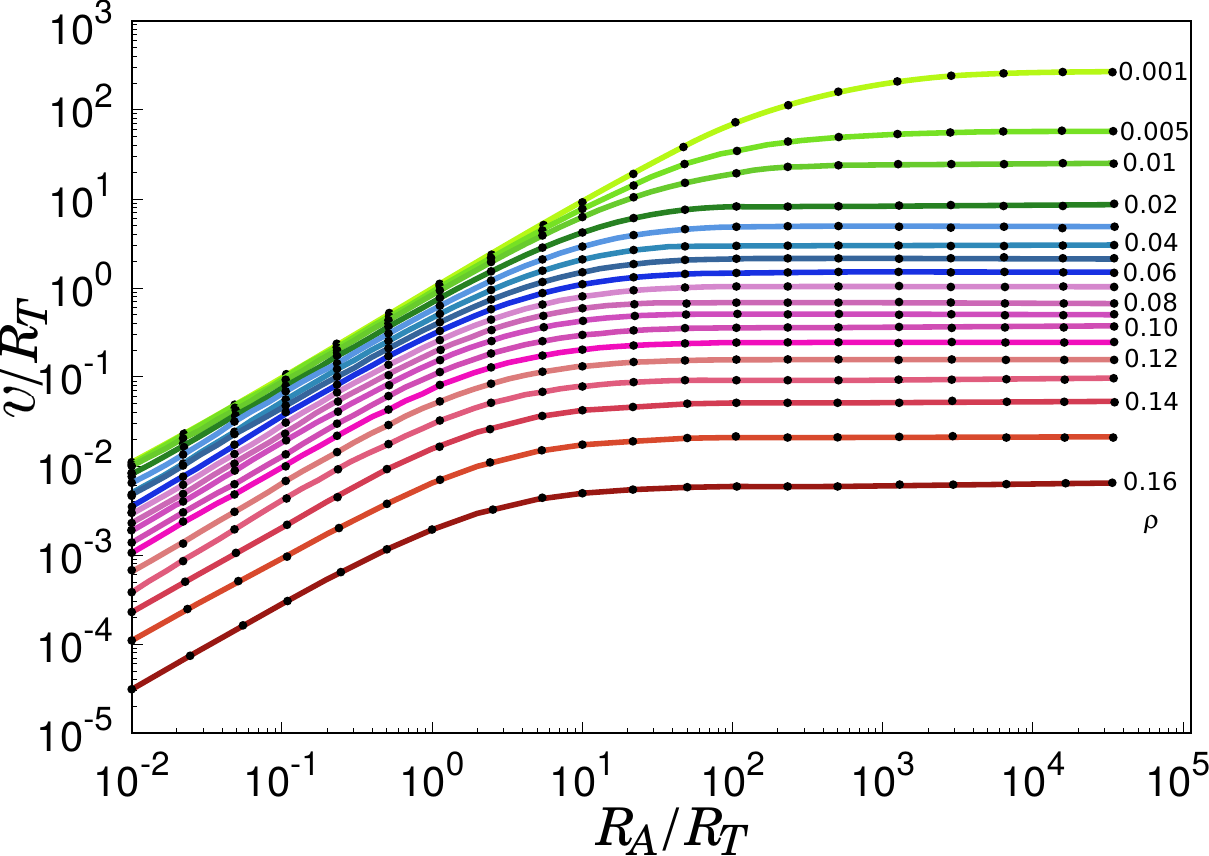}
\caption{The drift velocity of a non-rotating active tracer. Measuring the correlators of Eqs.~(\ref{eq:C+}-\ref{eq:C-}) from numerical simulation and plugging them in Eq.~(\ref{eqn:v_x_1}), lines, perfectly agrees with direct measurement of the drift velocity in simulation, points. Numbers next to each line indicate density values.}
\label{fig:v-ne-corr}
\end{figure}

We note that Eq.~(\ref{eqn:v_x_1}) is exact.  However, it requires the high-order non-equilibrium density correlations which appear in Eqs.~(\ref{eq:C+}-\ref{eq:C-}). Even without a theoretical framework for analytically calculating these correlations, we may obtain them from numerical simulations. Figure~\ref{fig:v-ne-corr} shows the perfect agreement between the direct measurement of the drift velocity $v$ in numerical simulations, and the evaluation of Eq.~(\ref{eqn:v_x_1}) using the correlations of Eqs.~(\ref{eq:C+}-\ref{eq:C-}) obtained in the same numerical simulations. Note that these are non-equilibrium simulations which include the active tracer in them. Below we will see which parts of the behavior that we observe here can be obtained using only equilibrium properties of the passive system.

\subsection{Tracer dynamics for low $R_A$}

Before proceeding, we note that if we ignore correlations, we may obtain for the low-density limit a mean-field approximation. To this end, we set in Eqs.~(\ref{eq:C+}-\ref{eq:C-}), $\la \eta_i \ra = \rho$, $\la \eta_i \eta_j \ra = \rho^2$, and $\la \eta_i \eta_j \eta_k \ra = \rho^3$. Thus we may write
\bea
{\cal{C}_+} = {\cal{C}_-} = {\cal{C}}_\textrm{MF} =  1 - 5 \rho + 5 \rho^2 - \rho^3 , \label{eq:CMF}
\eea
which yields the following mean-field drift velocity, $v_\textrm{MF} = R_A {\cal{C}}_\textrm{MF}$.

As seen in Fig.~\ref{fig:v-ne-corr}, at low $R_A$, by linear response the drift velocity $v$ is linearly proportional to $R_A$, and we may define the mobility as
\bea
\mu = \lim_{R_A \rightarrow 0} \frac{v}{R_A} .
\eea
We first note that the mean-field mobility is equal to $\mu_\textrm{MF} = {\cal{C}}_\textrm{MF}$, which is given in Eq.~(\ref{eq:CMF}) above.

\begin{figure}[t]
\includegraphics[width=\columnwidth]{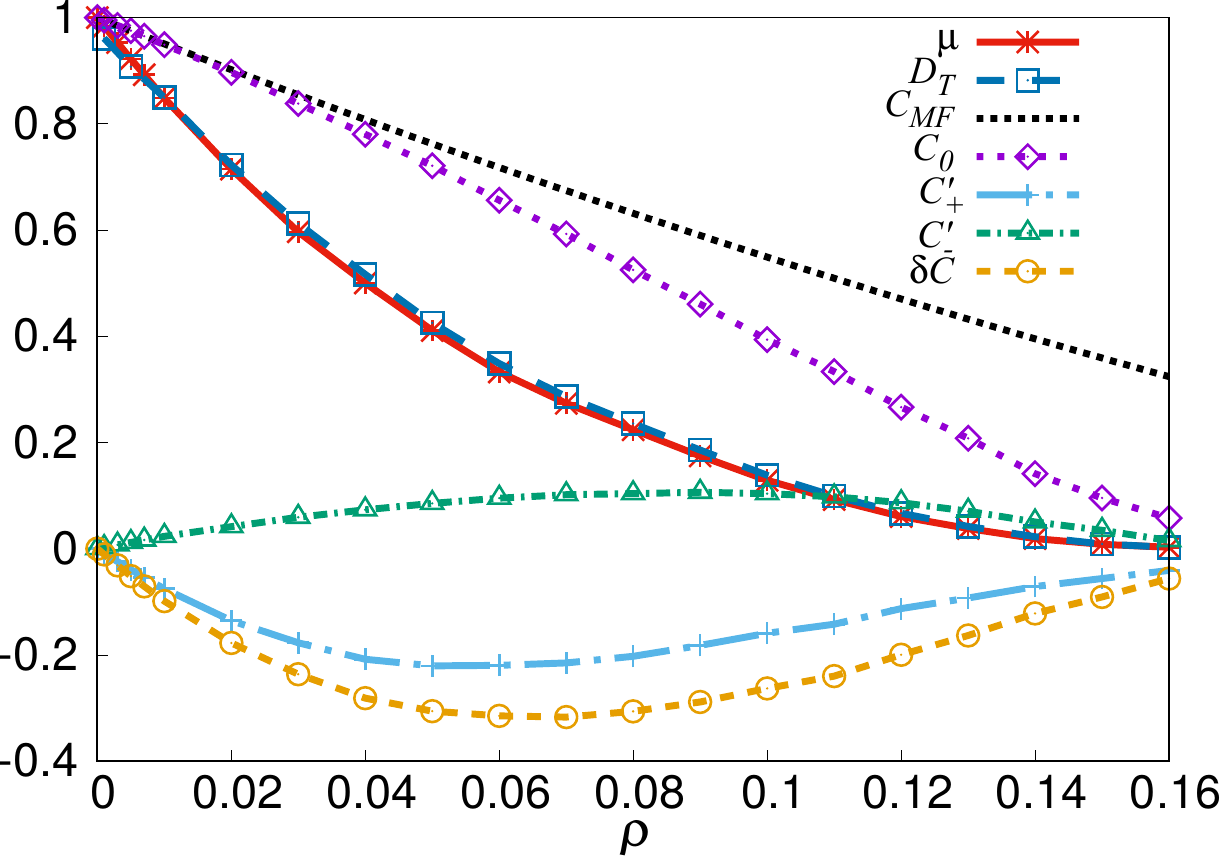}
\caption{Mobility and its different components vs density. Mobility $\mu$ perfectly agrees with the diffusion coefficient $D_T$. Mean-field result ${\cal{C}}_\textrm{MF}$ agrees with ${\cal{C}}_0$ at low densities. However even there, the additional terms in Eq.~(\ref{eq:mobility-expand}), namely $\delta {\cal{C}} = {\cal{C}'_{+}} -  {\cal{C}'_{-}}$ are significant.}
\label{fig:mobility}
\end{figure}

Now, given Eq.~(\ref{eqn:v_x_1}), we expand  ${\cal{C}_{\pm}}$ to linear order in $R_A$, 
\bea
{\cal{C}_{\pm}}(\rho,R_A) = {\cal{C}}_0 (\rho) + {\cal{C}'_{\pm}} (\rho) R_A , 
\eea
where prime indicates derivative with respect to $R_A$ at $R_A=0$. Thus
\bea
\mu = {\cal{C}}_0 + R_T \delta {\cal{C}} ,
\label{eq:mobility-expand}
\eea
where $\delta {\cal{C}} \equiv {\cal{C}'_+} - {\cal{C}'_-}$. That is, the mobility depends not only on the equilibrium value ${\cal{C}}_0$, but also on the forward-backward asymmetry encoded in $\delta {\cal{C}}$. Figure~\ref{fig:mobility} first shows that the Stokes-Einstein relations hold, namely the mobility exactly coincides with the self-diffusion coefficient measured from the long-time mean-squared displacement in a passive system, $\la \Delta r^2 \ra = 4 D_T t$. Secondly, the figure shows that the mean-field mobility ${\cal{C}}_\textrm{MF}$, Eq.~(\ref{eq:CMF}) describes well the low-densities behavior of the equilibrium correlator ${\cal{C}}_0$. Finally, and most importantly, Fig.~\ref{fig:mobility} shows how $\delta {\cal{C}}$ causes the actual mobility data to substantially deviate from ${\cal{C}}_0$, which encodes only equilibrium properties. We can say that ${\cal{C}}'_+<0$ since due to the active motion, with increasing activity neighboring sites in the forward direction are more likely to be occupied. Similarly ${\cal{C}}'_->0$ because in the backward direction, with increasing activity sites are more likely to be vacant. However, at this point we do not have even a low-density approximate theory for ${\cal{C}'_{+}}$ or ${\cal{C}'_{-}}$. 

\begin{figure}[t]
\includegraphics[width=\columnwidth]{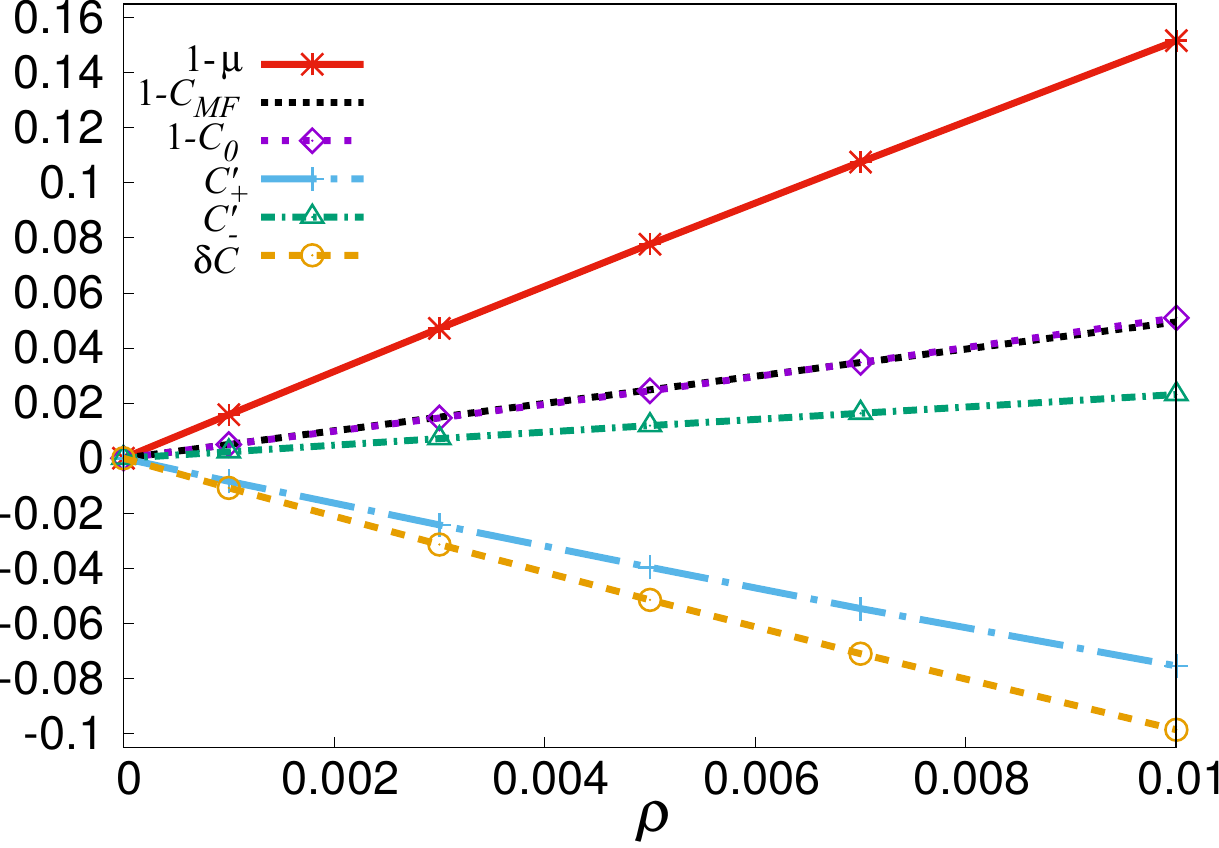}
\caption{Low-density behavior of the different components of the mobility. The non-equilibrium part $\delta {\cal{C}}$ has a lager contribution than the equilibrium part ${\cal{C}}_0$.}
\label{fig:mobility_zoom}
\end{figure}

To emphasize the importance of non-equilibrium information even at low activity and low density, we show in Fig.~\ref{fig:mobility_zoom} the extremely low-density behavior of each of the lines from Fig.~\ref{fig:mobility}. As expected, all exhibit linear dependence on density, with ${\cal{C}}_0 \approx {\cal{C}}_\textrm{MF} \approx 1 - 5 \rho$, ${\cal{C}}_+ \approx - 8 \rho$,  and ${\cal{C}}_- \approx 2 \rho$. Thus $\delta {\cal{C}} \approx - 10 \rho$, and $\mu \approx 1 - 15 \rho$. This numerically demonstrates the huge effect that non-equilibrium behavior has on the near-equilibrium behavior of the system. The equilibrium properties of the system are far from being able to describe the small deviations from equilibrium encoded in the mobility.

\subsection{Tracer dynamics for high $R_A$}

At large activity, the drift velocity $v$ of the tracer reaches a density-dependent asymptotic value $v_\infty$, as seen in Fig.~\ref{fig:v-ne-corr}. For extremely large $R_A$, once the active tracer can move in its active direction, it will immediately move. Hence it spends almost all of the time waiting for a thermal move to free it, and so to enable the active motion to resume. Here we develop a theoretical description for this process that will be valid in the low-density limit. At large $R_A$, we assume that the tracer moves rapidly into a region which has not been perturbed by its motion. The motion of the tracer involves instantaneous active flights between obstructions, along with periods spent waiting for an obstruction to be removed by a thermal move.

\begin{figure}[t]
\includegraphics[width=\columnwidth]{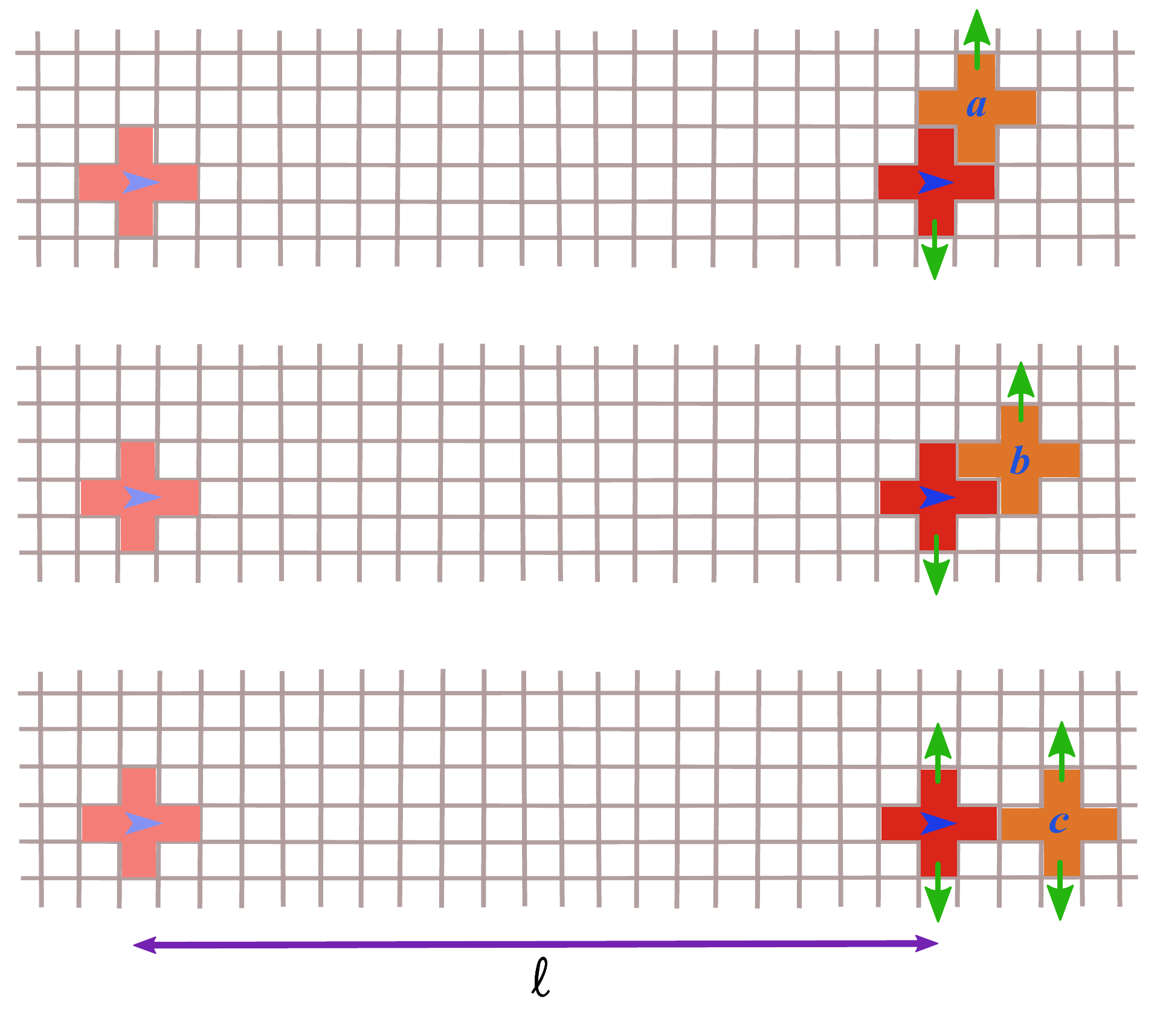}
\caption{Possible encounters of an active tracer (red) with a passive particle (orange) in the low-density, high-activity limit. The faded red cross indicates the preceding position at which the active tracer encountered a passive particle. The green arrows indicate the moves of the active tracer or of the passive particle that would lead to unblocking, and will enable the active tracer to continue moving along its active direction.} 
\label{fig:moves-highRA}
\end{figure}

We can describe this process as a continuous-time random-walk with steps of length $\ell_i$, along which the tracer moves in its active direction without meeting any passive particles that block its motion, see Fig.~\ref{fig:moves-highRA}. At the end of each flight the tracer has to wait for a time $\tau_i$ before it can begin its next flight. Assuming these two events are uncorrelated, the asymptotic drift velocity is given by the ratio between the average distance $\la \ell \ra$ traveled in each step, and the average waiting time $\la \tau \ra$ at the end of each step.

We first compute the average distance between obstructions $\la \ell \ra$, assuming the system around the active tracer to be in equilibrium. Therefore $\la \ell \ra$ depends only on the equilibrium density $\rho$. For low densities, at each new site the active tracer reaches, the probability that it will be able to continue to one more site is equal to the equilibrium value ${\cal{C}}_0(\rho)$, discussed above. So the probability of encountering an obstruction is $1- \mathcal{C}_0 (\rho)$. Treating obstructions as a Poisson process with density $1- \mathcal{C}_0(\rho)$, the distribution of gaps between them is given by $p(\ell) = \left(1- \mathcal{C}_0\right) \exp[ -\left( 1-\mathcal{C}_0 \right) \ell]$. Thus the average distance that the active tracer travels until reaching a passive particle that blocks it is equal to
\bea
\la \ell \ra = \frac{1}{1-{\cal{C}}_0(\rho)} . \label{eq:ell_av}
\eea

Next, we compute the average waiting time $\langle \tau \rangle$ at the end of each active flight. Once the active tracer meets a passive particle, it waits for a duration $\tau$ until it can continue its flight along its active direction. Therefore the active tracer must wait until the passive particle moves out of its way, or alternatively until the active particle itself moves laterally and the passive particle no longer blocks its motion along the active direction. For the active particle to be unblocked, the five sites $a$, $b$, $c$, $d$, and $e$ in front of it should be vacant (see Fig.~\ref{fig:active-nonrot}). In the low-density limit we may consider only single-particle blocking mechanisms. We consider a blocking particle to be in each one of the five blocking sites, $a$, $b$, $c$, $d$, or $e$. Each of these cases can have  different average waiting times, however symmetry dictates $\la \tau_a \ra = \la \tau_e \ra$ and $\la \tau_b \ra = \la \tau_d \ra$. Since these cases all involve single particle obstructions, each occurs with an equal probability that depends only on $\rho$. Therefore we can write the average waiting time for unblocking as 
\bea
\la \tau \ra = \frac{2}{5} \la \tau_a \ra + \frac{2}{5} \la \tau_b \ra + \frac{1}{5} \la \tau_c \ra . \label{eq:tau_av}
\eea

When the active tracer meets a blocking particle at site $a$, it has to wait until either the active particle thermally moves downwards, or the blocking particle thermally moves upwards, see Fig.~\ref{fig:moves-highRA}a. Each one of these processes occurs at rate $R_T$, thus the average time until one of them occurs is equal to $\la \tau_a \ra = \frac{1}{2R_T}$. Note that the particle at $a$ can also move forward (to the right in the figure). However, then the active tracer would immediately move forward and would still be blocked by this passive particle. We do not consider that as an unblocking event, because it merely increases $\ell$ by one. Since $\ell$ is much larger than one, this move has a negligible effect on $\la \ell \ra$.

If the blocking particle is at site $b$, due to the same argument we ignore its motion in the forward direction, and only consider the lateral motion. For the active tracer to overcome the blocking by this particle, these two particles should make two consecutive thermal moves in the lateral direction - the thermal particle upwards and the active particle downwards, see Fig.~\ref{fig:moves-highRA}b. Each single move occurs at rate $R_T$, thus the average time until either one moves one step laterally is $\frac{1}{2R_T}$, and the total time until two such lateral moves occur is equal to $\la \tau_b \ra = \frac{1}{R_T}$.

For site $c$ the first move can be one of four, see Fig.~\ref{fig:moves-highRA}c, thus takes an average time $\frac{1}{4R_T}$. It should then be followed by two more moves, that each has two options, similarly to the sequence of unblocking after meeting a particle at site $b$. This eventually leads to $\la \tau_c \ra = \frac{1}{4R_T} + \frac{2}{2R_T} = \frac{5}{4R_T}$. Plugging $\la \tau_a \ra$, $\la \tau_b \ra$, and $\la \tau_c \ra$ in Eqs.~(\ref{eq:ell_av}-\ref{eq:tau_av}) leads to the following approximation for the asymptotic drift velocity,
\bea
v_\infty = \frac{\la \ell \ra}{\la \tau \ra} = \frac{20 R_T}{17 \left(1 - {\cal{C}}_0 \right)} . \label{eq:v_asymp_theory}
\eea

\begin{figure}[t]
\includegraphics[width=\columnwidth]{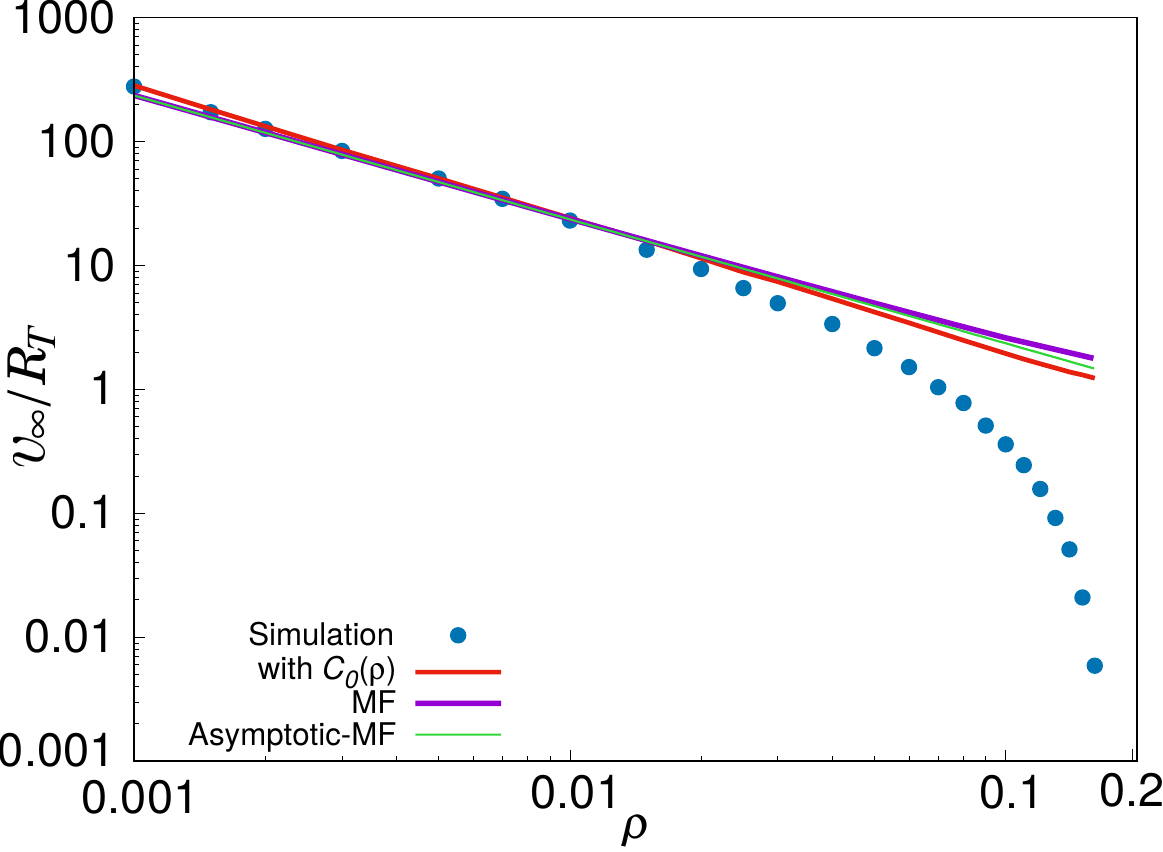}
\caption{Asymptotic, infinite $R_A$ drift velocity of non-rotating active tracer. Theory (lines) describes the low-density behavior of the simulation results (blue dots). In this low-density range, using the numerically-obtained correlation ${\cal{C}}_0(\rho)$, Eq.~(\ref{eq:v_asymp_theory}) does not differ that much compared to using the mean-field result, Eq. (\ref{eq:v_asymp_MF}), or its asymptotic behavior, Eq.~(\ref{eq:v_asymp_MF0}).}
\label{fig:v_asymptotic}
\end{figure}

Figure~\ref{fig:v_asymptotic} shows the agreement of this expression with the numerical results. We emphasize that in the low-activity limit we could not obtain a result for the drift velocity purely in terms of properties of the equilibrium system. However, in the high-activity limit studied here we can. We can further simplify Eq.~(\ref{eq:v_asymp_theory}) by substituting the mean-field expression ${\cal{C}}_\textrm{MF}$ from Eq.~(\ref{eq:CMF}) to get
\bea
v_\infty^{MF} = \frac{20 R_T}{17 \left(5\rho - 5\rho^2 + \rho^3 \right)} . \label{eq:v_asymp_MF}
\eea
In the low-density limit this may be further approximated to the following asymptotic form
\bea
v_\infty^{AMF} = \frac{4 R_T}{17 \rho} . \label{eq:v_asymp_MF0}
\eea
Since the theory developed here, and leading to Eq.~(\ref{eq:v_asymp_theory}) was valid for low densities in the first place, the agreement with numerical simulations shown in Fig.~\ref{fig:v_asymptotic} is only at low densities, where the differences between the different expressions (\ref{eq:v_asymp_theory},\ref{eq:v_asymp_MF},\ref{eq:v_asymp_MF0}) are very small.

\section{Rotation without locking} \label{sec:no-lock}

\subsection{Zero-density limit}

We will begin the derivation with the zero-density limit of a single isolated tracer, and will later introduce the effect of density. In this limit, a passive tracer moves at rate $R_T$ to each one of its four neighboring sites. Thus the stochastic evolution of its position may be written as
\begin{eqnarray}
x(t+dt) = 
\begin{dcases}
 \underline{\textrm{value}}:  &  \underline{\textrm{probability:}} \\
 x(t) + 1 & R_Tdt \\
 x(t) - 1 & R_Tdt \\
 x(t)     & 1 - 2R_Tdt .
\end{dcases}
\end{eqnarray}
Squaring and averaging over the stochasticity leads to
\bea
\la x^2(t+dt) \ra = \la x^2(t) \ra + 2 R_T dt ,
\eea
thus $\la x^2(t) \ra = 2R_T t$, and by symmetry $\la r^2(t) \ra = \la x^2(t) + y^2(t) \ra = 4 R_T t$. Thus the diffusion coefficient of this passive tracer equals $D_T \equiv \frac{\la r^2(t) \ra}{4t} = R_T$.

For an active tracer that rotates at rate $R_R$ without locking, we divide time into intervals $\Delta t_i$ between consecutive rotations. During each time interval, in the direction perpendicular to the active direction, the active tracer performs a random walk solely due to passive moves, thus $\la \Delta r_\perp^2(\Delta t) \ra = 2 R_T \Delta t$. For the parallel direction, in the present zero-density limit we employ Eq.~(\ref{evolution_eqns}) with $p_+ = R_T+R_A$ and $p_- = R_T$. Averaging we get a drift $\la \Delta r_\parallel (t) \ra= R_A t$. Squaring Eq.~(\ref{evolution_eqns}) and averaging we obtain
\bea
\la r_\parallel^2(t+dt) \ra = \la r_\parallel^2(t) \ra + (2R_T+R_A)dt +  2 R_A \la r_\parallel(t) \ra dt .
\eea
dividing by $dt$, substituting the drift expression obtained above, and integrating over a time interval $\Delta t$ between rotations, we get 
\bea
\la \Delta r_\parallel^2 (\Delta t) \ra = (2R_T+R_A) \Delta t + R_A^2 \Delta t^2 . 
\eea

We now average over the interval duration. Rotation to each one of the two directions occurs at rate $R_R$, thus the time intervals between rotations have a Poisson distribution $P(\Delta t) = 2 R_R \exp(- 2 R_R \Delta t)$. We may therefore write $\la \Delta t \ra = \frac{1}{2 R_R}$ and $\la \Delta t^2 \ra = \frac{1}{2 R_R^2}$, thus $\la \Delta t^2 \ra = \frac{\la \Delta t \ra}{R_R}$. Hence 
\bea
\la \Delta r^2 \ra &=& \la \Delta r_\parallel^2 \ra + \la \Delta r_\perp^2 \ra = (4R_T+R_A) \la \Delta t \ra + R_A^2 \la \Delta t^2 \ra \nonumber \\ 
&=& \left(4R_T + R_A + \frac{R_A^2}{R_R} \right) \la \Delta t \ra ,
\eea
and we obtain the following expression for the diffusion coefficient,
\bea
D_0 = \frac{\la \Delta r^2 \ra}{4 \la \Delta t \ra} = R_T + \frac{R_A}{4} + \frac{R_A^2}{4R_R} , \label{eq:D0}
\eea
where the subscript $0$ indicates the zero-density limit assumed above. We will discuss the different terms after extending this to finite densities.

\subsection{Including finite-density effects}

For finite density, the attempt rates for motion are given by $R_T$ and $R_A$, but due to the occupation of neighboring sites, not all attempts succeed. The long-time behavior of a passive particle is diffusive with the passive diffusion coefficient $D_T(\rho)$ discussed above (see Fig.~\ref{fig:mobility}). Thus we assume that our active tracer has probabilities $D_T(\rho)$ per unit time to perform thermal moves to each one of its four nearest neighbors. Similarly, during an interval between rotations, the active moves yield the drift velocity $v(\rho,R_A)$ discussed above (see Fig.~\ref{fig:v-ne-corr}), and we hence assume that the tracer has probability $v(\rho,R_A)$ per unit time to move in its active direction. Under this assumption that ignores temporal correlations in the success probabilities of attempted moves, we may replace $R_T$ in the zero-density derivation, Eq.~(\ref{eq:D0}) above by $D_T(\rho)$ and $R_A$ by $v(\rho,R_A)$, leading to
\bea
D = D_T(\rho) + \frac{v(\rho,R_A)}{4} + \frac{v^2(\rho,R_A)}{4R_R} . \label{eq_D_no_lock}
\eea

We assumed that the passive and active motions are uncorrelated random processes, thus the total diffusion coefficient we obtained is equal to the diffusion coefficient in the passive case, plus the diffusion coefficient resulting from the active process. The last term is similar to what we would get for an active Brownian particle~\cite{Fily2014}. However, the second term, which is linear in the drift velocity is a result of the discrete nature of the motion on the lattice~\cite{Benichou2013A}. One way to understand this term is to consider the fast rotation limit, $R_R \gg R_A$. In that limit, each time an active move is attempted the active direction has been completely randomized, and the added rate $R_A$ may be thought to be uniformly distributed between the rates of moving in all four directions, thus the tracer undergoes passive motion with an effective thermal rate $R_T+\frac{R_A}{4}$.

\begin{figure}[t]
\includegraphics[width=\columnwidth]{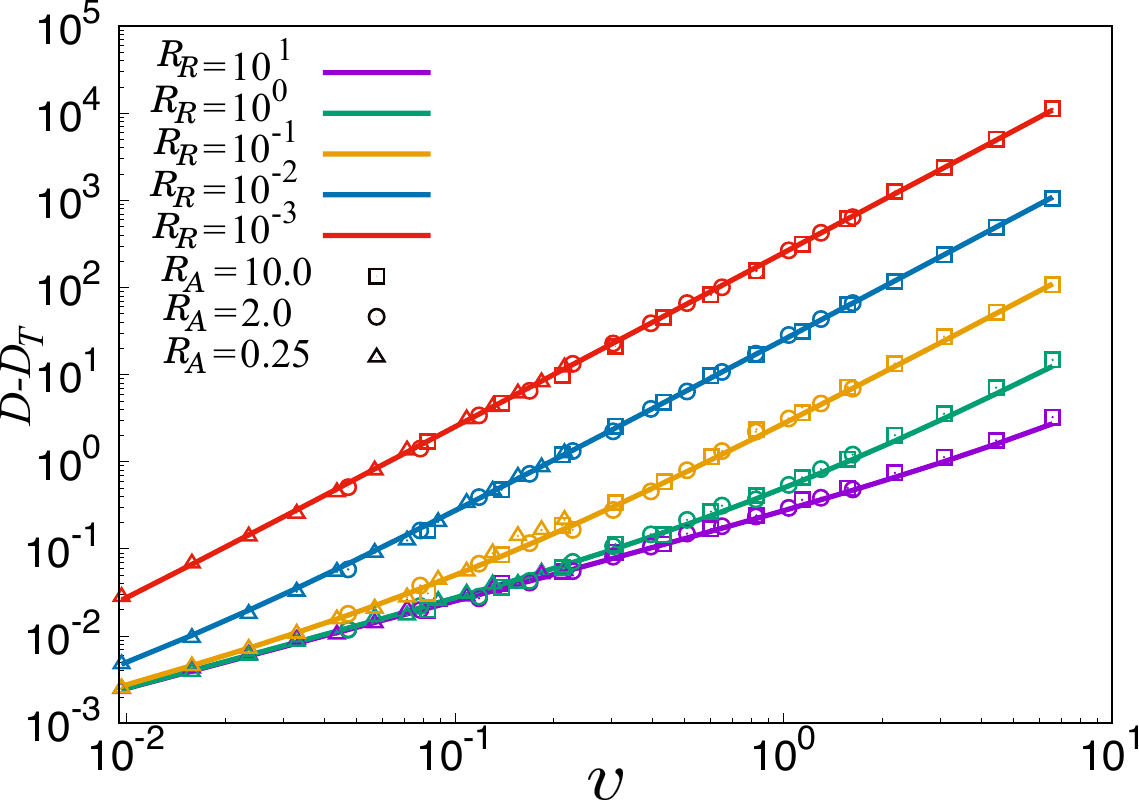}
\caption{Active part of the diffusion coefficient of an active tracer that rotates without rotational locking, plotted vs. the drift velocity, as obtained from simulations without rotation. The solids lines represent $\frac{v}{4} + \frac{v^2}{4R_R}$, as predicted by Eq.~(\ref{eq_D_no_lock}).}
\label{fig:check_quad}
\end{figure}

Note that in the derivation of Eq.~(\ref{eq_D_no_lock}) we assumed that the active tracer moves at velocity $v(\rho,R_A)$ between rotations. In practice this velocity is obtained only after some time, and this derivation should be valid only for low enough rotation rate. To test this, we show in Fig.~\ref{fig:check_quad}, $D-D_T$ vs $v$ for multiple $\rho$ and $R_A$ values, where each color corresponds to a different value of $R_R$, as indicated in the legend. The figure shows nice data collapse and agreement with Eq.~(\ref{eq_D_no_lock}) even at high $R_R$.

\section{Rotational Locking} \label{sec:lock}

\begin{figure}[t]
\includegraphics[width=\columnwidth]{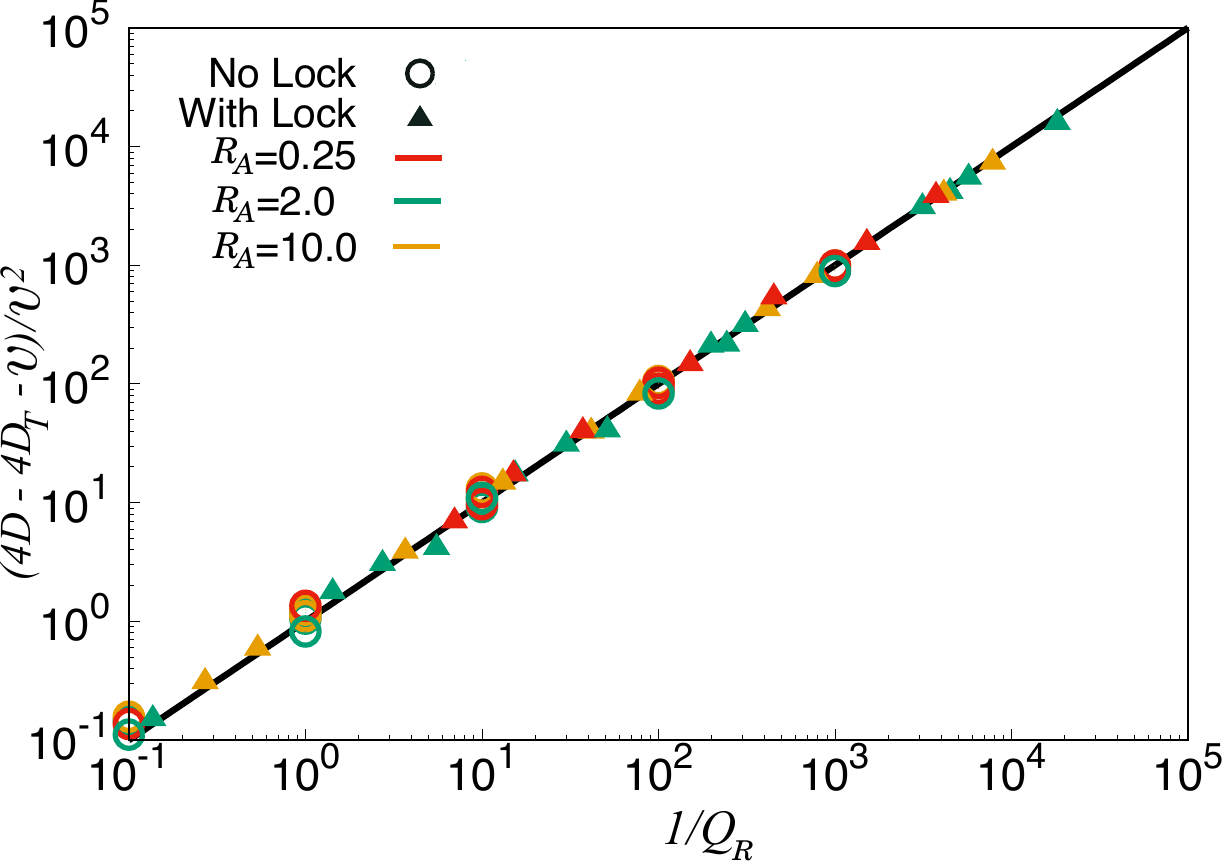}
\caption{Agreement with Eq.~(\ref{eq_D_with_lock}) of data from simulations with (solid symbols) and without (open symbols) rotational locking. Data collapses to the identity line, while all three parameters, $\rho$, $R_T$ and $R_R$ were varied for an arbitrary choice of $R_T=1$.}
\label{fig:QR-linear-scatter}
\end{figure}

When rotational locking is included, with increasing density not all rotation attempts succeed, thus the actual rotation rate $Q_R$ is slower than the rotation attempt rate $R_R$, and the diffusion coefficient is smaller, in agreement with Eq.~(\ref{eq_D_no_lock}). We now show how this argument may yield also a quantitative prediction. Namely, we measure the actual rotation rate $Q_R$ in simulations with rotational locking, and assume that we may generalize Eq.~(\ref{eq_D_no_lock}) to include $Q_R$ instead of $R_R$
\bea
D = D_T(\rho) + \frac{v(\rho,R_A)}{4} + \frac{v^2(\rho,R_A)}{4Q_R} , \label{eq_D_with_lock}
\eea
where with locking $Q_R<R_R$, while without locking $Q_R=R_R$. Figure~\ref{fig:QR-linear-scatter} shows that when plotted vs. $Q_R$, results with and without locking perfectly agree. It would be interesting to develop a theory for the rotation acceptance rate $Q_R/R_R$. We could obtain numerical results only for densities $\rho \le 0.13$, since at higher densities the dynamics are extremely slow, both due to the low drift velocity and due to rotational locking, which dramatically slows down the eventual rotation rate. Nonetheless, we expect that even as the density increases, Eq.~(\ref{eq_D_with_lock}) should describe the long-time diffusive behavior.

\section{Discussion}

We consider the lattice-gas model with exclusion up to third-order neighbors on the square lattice. Due to this exclusion, the particles in this model are equivalent to hard, cross-shaped pentamers. This model has an equilibrium first-order phase transition with coexistence between fluid at density $\rho_m \approx 0.16$ and crystal at density $\rho_x \approx 0.19$, and exhibits a glass transition at $\rho_g \approx 0.17$. We added activity to this model by assigning an active direction of self propulsion to the particles. With time, this direction may undergo rotational diffusion. The cross shape of the particles enabled us to naturally introduce a locking mechanism that prevents rotation due to the presence of neighboring particles. Our ultimate goal is to use this model to study the interplay between activity and jamming. This will advance the understanding of the cooperative phenomena that govern closely packed active matter systems.

In this paper we focused on the case of a single active tracer in a bath of thermal particles. At long times, this tracer undergoes diffusive motion, and for the case without rotational locking, we could write its self-diffusion coefficient in terms of i) the diffusion coefficient $D_T(\rho)$ of a passive particle, ii) the drift velocity $v(R_A,\rho)$ of a non-rotating active tracer, and iii) the rotation rate $R_R$. Remarkably, when rotational locking is introduced, it merely reduces the rate of successful rotations, and when the rate $Q_R$ at which rotations occur is used instead of the rate $R_R$ of rotation attempts, our theoretical prediction perfectly matches the results of our numerical simulations.

Interestingly, we showed that the properties of the passive system are not enough to fully describe the drift velocity of a non-rotating active tracer, not even in the low-activity and low-density limit. Here, we expect the system to be close to equilibrium, yet we could not describe the dynamics of the tracer only using knowledge of the equilibrium properties of the system. Thus, even in the low-density limit, where equilibrium correlations vanish, the mean-field approximation does not work. It is interesting to note that a similar phenomena has recently been studied in the context of bulk diffusion in lattice-gas models~\cite{Arita2014, Teomi2017, Arita2017, Teomi2019}. Nonetheless, we developed a theory in terms of equilibrium correlators, which applies for low density and strong activity.

There are several interesting directions to explore. We can add a small but finite density of active tracers to the thermal lattice gas, in which case interactions between the active particles, mediated through the bath, can show interesting effects. Finally our study can also serve as the basis for the investigation of this model when all the particles are active. For such an all-active system we also expect to be able to understand the long-time diffusive behavior of a rotating particle in terms of its drift velocity during intervals between rotations. And we expect to be able to describe the effect of rotational locking as reducing the rotation rate.

\begin{acknowledgments}

We thank Gregory Bolshak, Abhishek Dhar, Arghya Das, Arnab Pal, Shlomi Reuveni, and Eial Teomy for helpful discussions. This work was partially supported by the Israel Science Foundation grant No. 968/16, and by a grant from the United States-Israel Binational Science Foundation. CM and BC have been supported by the Brandeis MRSEC through NSF-DMR 1420382.

\end{acknowledgments}


\begin{thebibliography}{99}

\bibitem{parry_2014} B. R. Parry, I. V. Surovtsev, M. T. Cabeen, C. S. O'Hern, E. R. Dufresne, and C. Jacobs-Wagner, Cell \textbf{156}, 183 (2014).

\bibitem{chaudhuri_2011} A. Chaudhuri, B. Bhattacharya, K. Gowrishankar, S. Mayor, and M. Rao, Proc. Natl. Acad. Sci. USA \textbf{108}, 14825 (2011).

\bibitem{angelini_2011} T. E. Angelini, E. Hannezo, X. Trepat, M. Marquez, J. J. Fredberg, and D. A. Weitz, Proc. Natl. Acad. Sci. USA \textbf{108}, 4714 (2010).

\bibitem{bi_2015} D. Bi, X. Yang, M. C. Marchetti, and M. L. Manning, Phys. Rev. X \textbf{6}, 021011 (2015).

\bibitem{chen_2012} X. Chen, X. Dong, A. Be’er, H. L. Swinney, and H. P. Zhang, Phys. Rev. Lett. \textbf{108}, 148101 (2012).

\bibitem{silverberg_2013} J. L. Silverberg, M. Bierbaum, J. P. Sethna, and I. Cohen
Phys. Rev. Lett. {\bf 110}, 228701 (2013).

\bibitem{fily_2012} Y. Fily, and M. C. Marchetti, Phys. Rev. Lett. \textbf{108}, 235702 (2012).

\bibitem{redner_2013} G. S. Redner, M. F. Hagan, and A. Baskaran, Phys. Rev. Lett. \textbf{110}, 055701 (2013).

\bibitem{cates_2015} M. E. Cates, and J. Tailleur, Annu. Rev. Condens. Matter Phys. \textbf{6}, 219 (2015).

\bibitem{bialke_2012} J. Bialke, T. Speck, and H. L¨owen, Phys. Rev. Lett. \textbf{108}, 168301 (2012).

\bibitem{Liao_2018} Q. Liao, and N. Xu, Soft Matter \textbf{14}, 853 (2018). 

\bibitem{berthier_2014} L. Berthier, Phys. Rev. Lett. \textbf{112}, 220602 (2014).

\bibitem{rituparno_2016} R. Mandal, P. J. Bhuyan, M. Rao, and C. Dasgupta, Soft Matter \textbf{12}, 6268 (2016). 

\bibitem{Fily2014} Y. Fily, S. Henkes, and M. C. Marchetti, Soft Matter \textbf{10}, 2132 (2014).

\bibitem{levis_2014} D. Levis, and L. Berthier, Phys. Rev. E. \textbf{89}, 062301 (2014).  

\bibitem{redner_2013_b} G. S. Redner, M. F. Hagan, and A. Baskaran, Phys. Rev. E \textbf{88}, 012305 (2013). 

\bibitem{illien_2013} P. Illien, O. B\'enichou, C. Mejía-Monasterio, G. Oshanin, and R. Voituriez, Phys. Rev. Lett. \textbf{111}, 038102 (2013).

\bibitem{Benichou2013} O. B\'enichou, P. Illien, G. Oshanin, and R. Voituriez, Phys. Rev. E \textbf{87}, 032164 (2013). 

\bibitem{Benichou2014} O. B\'enichou, P. Illien, G. Oshanin, A. Sarracino, and R. Voituriez, Phys. Rev. Lett. \textbf{113}, 268002 (2014).

\bibitem{mallick_2011} T. Chou, K. Mallick, and R. Zia, Rep. Prog. Phys. \textbf{74}, 116601 (2011).

\bibitem{squires} T. M. Squires and T. G. Mason, Ann. Rev. Fluid Mech. \textbf{42}, 413 (2009).

\bibitem{Baxter1980} R. J. Baxter, J. Phys. A \textbf{13}, L61 (1980).

\bibitem{Dickman2012} R. Dickman, J. Chem. Phys. \textbf{136}, 174105 (2012).

\bibitem{Barnes2009} B. C. Barnes, D. W. Siderius, and L. D. Gelb, Langmuir \textbf{25}, 6702 (2009).

\bibitem{Fernandez2007} H. C. M. Fernandez, J. J. Arenzon, and Y. Levin, J. Chem. Phys. \textbf{126}, 114508 (2007).

\bibitem{Nath2014} T. Nath and R. Rajesh, Phys. Rev. E \textbf{90}, 012120 (2014).

\bibitem{Bellemans1996} A. Bellemans and J. Orban, Phys. Rev. Lett. \textbf{17}, 908 (1966).

\bibitem{Orban1982} J. Orban and D. Van Belle, J. Phys. A \textbf{15}, L501 (1982)

\bibitem{Eisenberg2005} E. Eisenberg and A. Baram, Europhys. Lett. \textbf{71}, 900 (2005).

\bibitem{Rotman2009} Z. Rotman and E. Eisenberg, Phys. Rev. E \textbf{80}, 060104 (2009).

\bibitem{Rotman2010} Z. Rotman and E. Eisenberg, Phys. Rev. Lett. \textbf{105}, 225503 (2010).

\bibitem{Eisenberg2000} E. Eisenberg and A. Baram, J. Phys. A: Math. Gen. \textbf{33}, 1729 (2000).

\bibitem{Ramola2012} K. Ramola and D. Dhar, Phys. Rev. E \textbf{86}, 031135 (2012). 

\bibitem{Barma1994} M. Barma and D. Dhar, Phys. Rev. Lett. \textbf{73}, 2135 (1994).

\bibitem{whitelam_2018} S. Whitelam, K. Klymko, and D. Mandal, J. Chem. Phys. \textbf{148}, 154902 (2018)

\bibitem{Benichou2013A} O. B\'enichou, K. Lindenberg, and G. Oshanin, Physica A \textbf{392}, 3909 (2013)

\bibitem{Arita2014} C. Arita, P. L. Krapivsky, and K. Mallick, Phys. Rev. E \textbf{90}, 052108 (2014).

\bibitem{Teomi2017} E. Teomy and Y. Shokef, Phys. Rev. E \textbf{95}, 022124 (2017).

\bibitem{Arita2017} C. Arita, P. L. Krapivsky, and K. Mallick, Phys. Rev. E \textbf{95}, 032121 (2017).

\bibitem{Teomi2019} E. Teomy, private communication.

\end{thebibliography}
\end{document}